\documentclass[prl,twocolumn,aps,showpacs]{revtex4}
\usepackage{graphicx}
\usepackage{color}
\usepackage{dcolumn}
\usepackage{multirow}
\usepackage{amsmath}
\usepackage{yhmath}
\usepackage[logonly]{trace}
\usepackage{amssymb}
\usepackage{subfigure}
\usepackage{latexsym}
\usepackage{mathrsfs}
\usepackage{wasysym}
\usepackage{ulem}

\begin{document}

\title{Impact of loss on the wave dynamics in photonic waveguide lattices}

\author{M. Golshani$^1$}
\email{golshanimojtaba@gmail.com}
\author{S. Weimann$^2$}
\author{Kh. Jafari$^3$}
\author{M. Khazaei Nezhad$^1$}
\author{A. Langari$^{1,4,5}$}
\author{A. R. Bahrampour$^1$}
\author{T. Eichelkraut$^2$}
\author{S. M. Mahdavi$^{1,6}$}
\author{A. Szameit$^2$}

\affiliation{$^1$Department of Physics, Sharif University of Technology, Tehran 11155-9161, Iran}

\affiliation{$^2$Institute of Applied Physics, Abbe Center of Photonics, Friedrich-Schiller-University Jena, Max-Wien-Platz 1, 07743 Jena, Germany}

\affiliation{$^3$Department of Physics and Institute for Plasma Research, Kharazmi University, Tehran 15614, Iran}

\affiliation{$^4$Center of Excellence in Complex Systems and Condensed Matter (CSCM), Sharif University of Technology, Tehran 145888-9694, Iran}

\affiliation{$^5$Max-Planck-Institut f\"ur Physik komplexer Systeme, 01187 Dresden, Germany}

\affiliation{$^6$Institute for Nanoscience and Nanotechnology, Sharif University of Technology, Tehran, Iran}


\begin{abstract}
We analyze the impact of loss in lattices of coupled optical waveguides and find that in such case, the hopping between adjacent waveguides is
necessarily complex. This results not only in a transition of the light spreading from ballistic to diffusive, but also in a new kind of diffraction
that is caused by loss dispersion. We prove our theoretical results with experimental observations.
\end{abstract}

\maketitle


Absorption is an intrinsic feature of photonic systems, arising due to the laws of causality \cite{BornWolf}. It results in decoherence and, hence,
in a considerable change in the dynamics of optical waves. However, it is generally agreed that in the particular case of homogeneous and isotropic
loss the impact on the amplitude distribution in the system vanishes, besides a global decay of the integrated power \cite{BornWolf}. A very
prominent photonic system is arrays of evanescently coupled waveguides \cite{LedererNature}, where a tailored absorption (or absorption/gain)
distribution is the basis for a multitude of unexpected physical phenomena, such as exceptional points \cite{Guo}, unusual beam dynamics
\cite{Makris-Musslimani}, spontaneous $\mathcal{PT}$-symmetry breaking \cite{ruter}, non-reciprocal Bloch oscillations \cite{BlochLonghi} and dynamic
localization \cite{DynamicsLonghi}, unidirectional cloaking \cite{Regensburger}, and even tachyonic transport \cite{Szameit}. Owing to the intuition
described above, if all lattice sites exhibit exactly the same absorption, its impact vanishes in the evolution equations of these systems. In a more
mathematical language, in this case absorption adds to the Hamiltonian as a pure diagonal matrix with identical elements, which can be removed by
normalization.

In our work we show that absorption in coupled waveguide systems does always impact the light dynamics, even if it is homogeneous and isotropic in all
lattice sites. Due to the imaginary part of the dielectric function (that describes the absorption) imaginary off-diagonal elements in the
Hamiltonian appear that cannot be removed by normalization, causing significant deviations in the light dynamics compared to the Hermitian case.
However, our theory holds for all Schr\"odinger type systems that can be mapped onto a tight binding lattice, e.g., paraxial waves in optics or mechanics as well as quantum dynamics in spin chains, population transfer in multi-level systems and graphene. Our theory supplements the knowledge about the influence of non-Hermiticity to all these systems in general including the effect of PT symmetry.

\begin{figure}[b]
\includegraphics[width=\columnwidth]{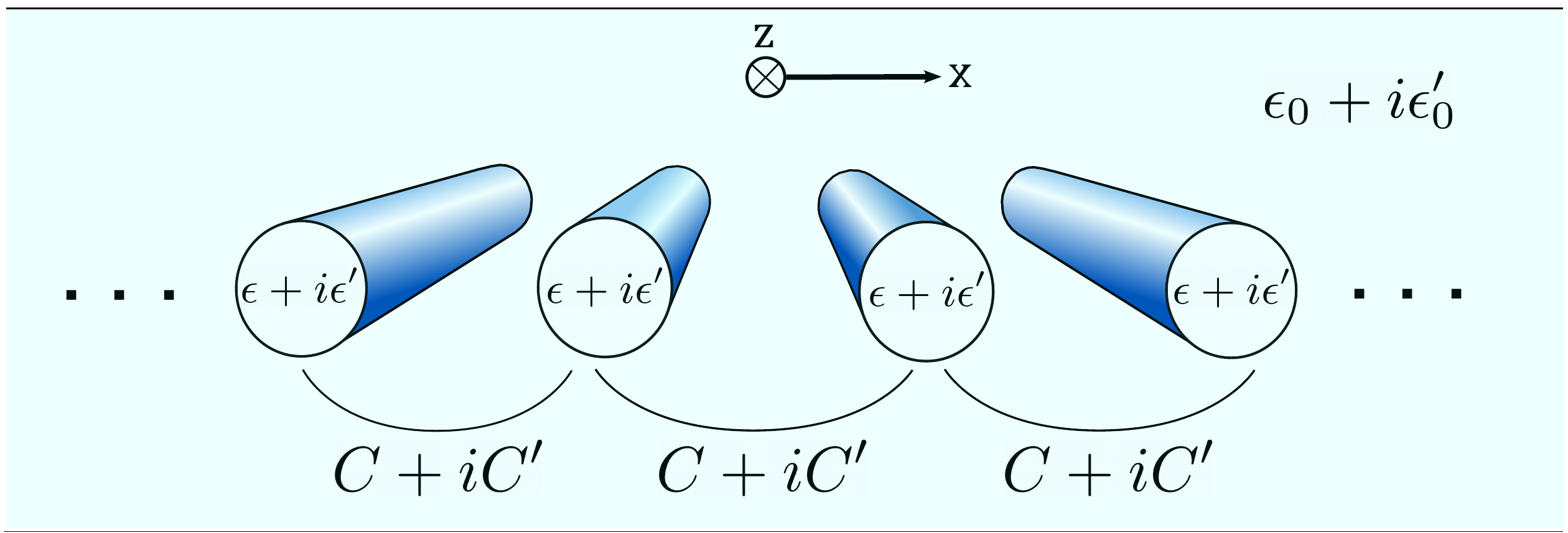}
\caption{(color online) One-dimensional array of identical absorbing optical waveguides. The complex relative electric permittivity of all waveguides is
${\epsilon} + i\epsilon {'}$, while the surrounding medium is fused silica with relative electric permittivity ${\epsilon _0} + i\epsilon {'_0}$.}
\label{Fig1}
\end{figure}

In order to study the impact of absorption in such systems, we consider a one-dimensional array of $N$ identical single mode optical waveguides with
width $2w$, inter-site spacing $d$, and the complex relative electric permittivity ${\epsilon} + i\epsilon {'}$ at the positions $x_n$ ($n=1,2,...,N$), which
is surrounded by a bulk material (with ${\epsilon _0} + i\epsilon {'_0}$). A sketch of this system is shown in Fig. \ref{Fig1}.

\par The dynamics of wave propagating through this system is governed by the Helmholtz wave equation
\begin{eqnarray}
\label{eqHelmholtz} \left ({\nabla ^2} + k_0^2\widetilde \varepsilon (x) \right )\psi (x,z) = 0\; ,
\end{eqnarray}
where $\psi(x,z)$ is the electric field amplitude, $k_0=\frac{\omega }{c}$ is the propagation constant in free space, and $\widetilde \varepsilon
(x)$ is relative electric permittivity profile of the system. The relative electric permittivity distribution of the entire structure can be written as a
sum of individual waveguide contributions, such that
\begin{eqnarray}
\label{eqEpsilonProfile} \widetilde \varepsilon (x) = {\epsilon _0} + i\epsilon {'_0} + \sum\limits_{n=1}^N {[(\epsilon  - {\epsilon _0}) +
i(\epsilon ' - \epsilon {'_0})]{\zeta _n}(x)} \; .
\end{eqnarray}
Here, we used ${\zeta _n}\left( x \right) = H\left( {x - {x_n} + w} \right) - H\left( {x - {x_n} - w} \right)$ (with $H\left( {x} \right)$ as the
Heaviside step function). In the tight-binding approximation, the full field $\psi (x,z)$ can be written as a superposition of individual waveguide
modes
\begin{eqnarray}
\label{eqCoupledModeApprox} \psi (x,z) = \sum\limits_{n=1}^N {{\phi _n}(z)u(x - {x_n}){e^{i\beta z}}} \; ,
\end{eqnarray}
where $\beta$ is the waveguide's propagation constant ($k_0 \sqrt{\epsilon_0} < \beta < k_0 \sqrt{\epsilon}$), whereas $u(x-{x_n})$ and ${\phi
_n}(z)$ represent the normalized transverse mode profile and the field amplitude in $n$th waveguide, respectively. After a somewhat lengthy but
straightforward calculation (see Supplemental Material for details) one obtains the coupled-mode equations for the light evolution in the non-Hermitian lattice
\begin{equation}\label{EqTightBinding}
 - i\frac{{d{{\phi}_n}}}{{dz}} = i \kappa {\phi}_n + (C + iC')\left({\phi}_{n+1} + {\phi}_{n-1}\right) \;.
\end{equation}
Here,
\begin{equation}\label{losscoefficient} \kappa  = \frac{k_0^2}{2\beta}
\left (\epsilon'_0 + (\epsilon ' - \epsilon'_0)\mathrm{tanh}\big(\frac{w}{\ell}\big) \right),
\end{equation}
is the loss coefficient ($\ell$ is the width of the eigenmode), and
\begin{eqnarray}
\label{coupling_realpart}
C &=& \frac{(\epsilon  - \epsilon _0)k_0^2}{2\beta}\frac{w}{\ell}\exp\big(-\frac{d}{\ell}\big),\\
\nonumber \\ \label{coupling_imagpart}
C' &=& \frac{(\epsilon'_0 - \epsilon')k_0^2}{\beta}\frac{d}{\ell }\exp\big(-\frac{d}{\ell}\big),
\end{eqnarray}
represents the real and imaginary part of the inter-site hopping rate, respectively. Note that the diagonal term $i \kappa {\phi}_n$ can be removed
by the normalization ${\phi}_n = E_ne^{-\kappa z}$, whereas the off-diagonal terms $i C' {\phi}_n$ cannot. It is therefore evident that for
\textit{any} absorption present in the waveguides the light dynamics will be affected. Interestingly, for a given absorption profile, one finds the
relation
\begin{equation}\label{ratiocoupling}
C' = \alpha C ~,
\end{equation}
between the real and the imaginary part of the inter-site hopping, with
\begin{equation}
\label{absorptiondiscrepancy} \alpha  = 2\frac{(\epsilon {'_0} - \epsilon ')}{(\epsilon  - \epsilon_0)}\frac{d}{w} ~,
\end{equation}
as the absorption discrepancy. Therefore, the imaginary part $C'$ is always in a fixed ratio to the real part $C$ of the hopping. Note that the
absoprtion discrepancy itself is proportional to the inter-site spacing $d$. We would like to note that the absorption discrepancy $\alpha$ vanishes
for $\epsilon'_0 \rightarrow \epsilon'$, i.e., when not only the absorption in the lattice is homogeneous, but the absorption in the entire system
(that is, in the lattice and the surrounding bulk material).

\begin{figure}[b]
\includegraphics[width=\columnwidth]{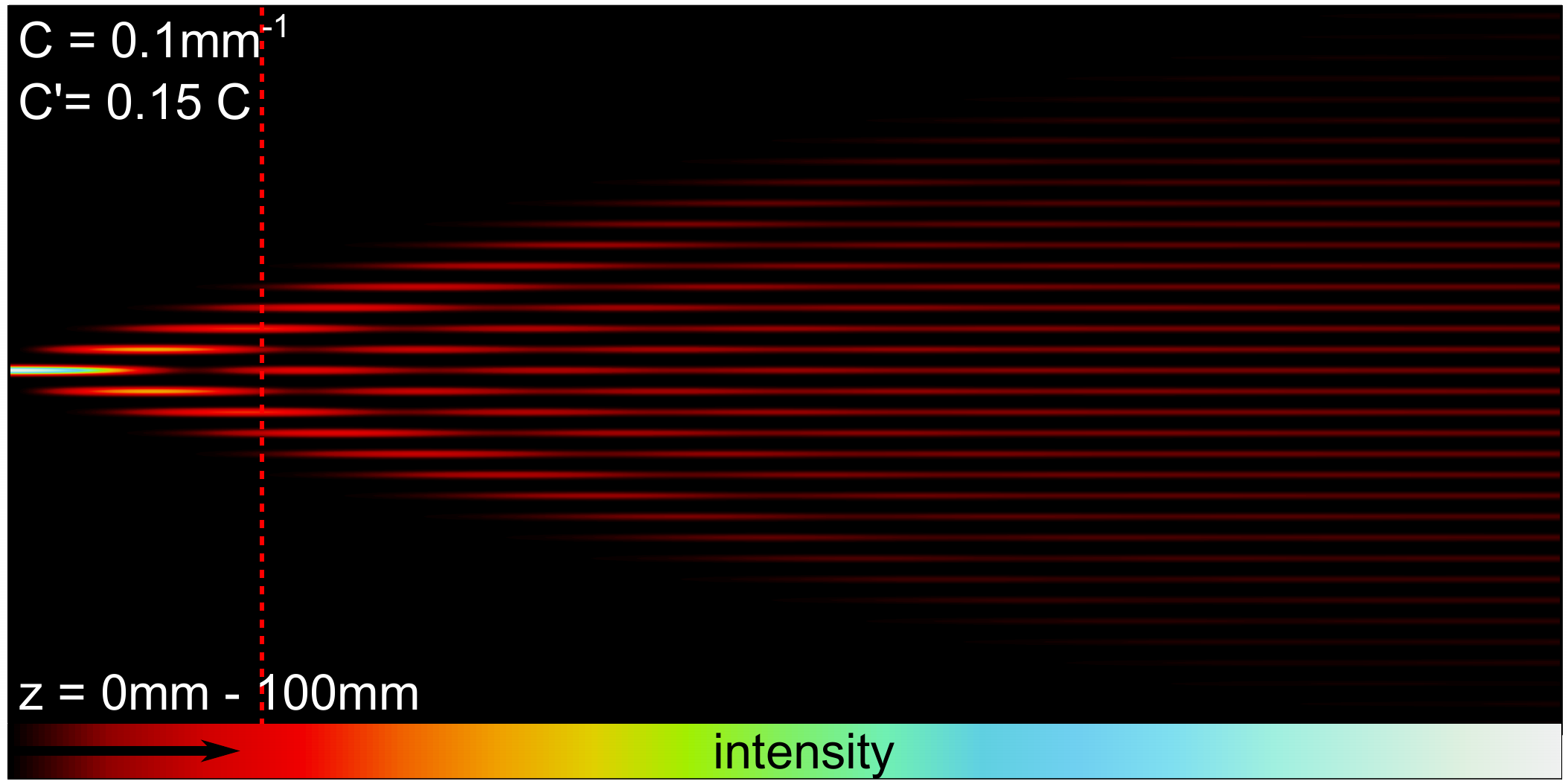}
\caption{(color online) Evolution in a waveguide array, where each waveguide exhibits the same loss, resulting in $\alpha = 0.15$. Clearly, after
$z_{crit}$ (red dashed line), the light spreading changes from ballistic to diffusive. The power is normalized to 1 at every $z$.} \label{Fig2}
\end{figure}

There are several important consequences arising from the appearance of an additional imaginary off-diagonal term in the Hamiltonian. First, we find
that, for any loss discrepancy (i.e., $\alpha\not = 0$), the light spreading is ballistic for distances $z\ll z_{\mathrm{crit}}$ with
\begin{equation} \label{z_crit}
  z_{\mathrm{crit}} =\frac{1}{4\alpha C} \; ,
\end{equation}
but slows down to diffusive for $z\gg z_{\mathrm{crit}}$ (see Fig. \ref{Fig2}). This can be seen by taking into account the Green's function of Eq.
\eqref{EqTightBinding}
\begin{equation}
  E_n (z) = i^n J_n(2(1+i\alpha)Cz) \; .
\end{equation}
The variance of this evolving wave packet is (see Supplemental Material for details on the calculation)
\begin{equation}\label{VarianceofWavepacket}
  \sigma^2 \left( z \right) = \left( \alpha  + \frac{1}{\alpha } \right)Cz\,\frac{I_1\left( 4\alpha Cz \right)}{I_0\left( 4\alpha Cz\right)} \; ,
\end{equation}
which can be approximated as
\begin{eqnarray}
  \sigma ^2\left( z \right)\mathop  \to \limits^{4\alpha Cz \ll 1} 2\left( 1 + \alpha^2 \right)C^2 z^2 \quad \mathrm{(ballistic)} ,\\[1ex]
 \sigma ^2\left( z \right)\mathop  \to \limits^{4\alpha Cz \gg 1} \left( \frac{1 + \alpha ^2}{\alpha} \right)Cz \quad \mathrm{(diffusive)} \; .
\end{eqnarray}
Hence, even for minimal loss decoherence effects impact the wave packet evolution, resulting eventually in a diffusive
spreading behavior for
sufficiently large propagation distances despite the fact that the lattice exhibits full translational symmetry.

\begin{figure}[b]
\includegraphics[width=0.95 \columnwidth]{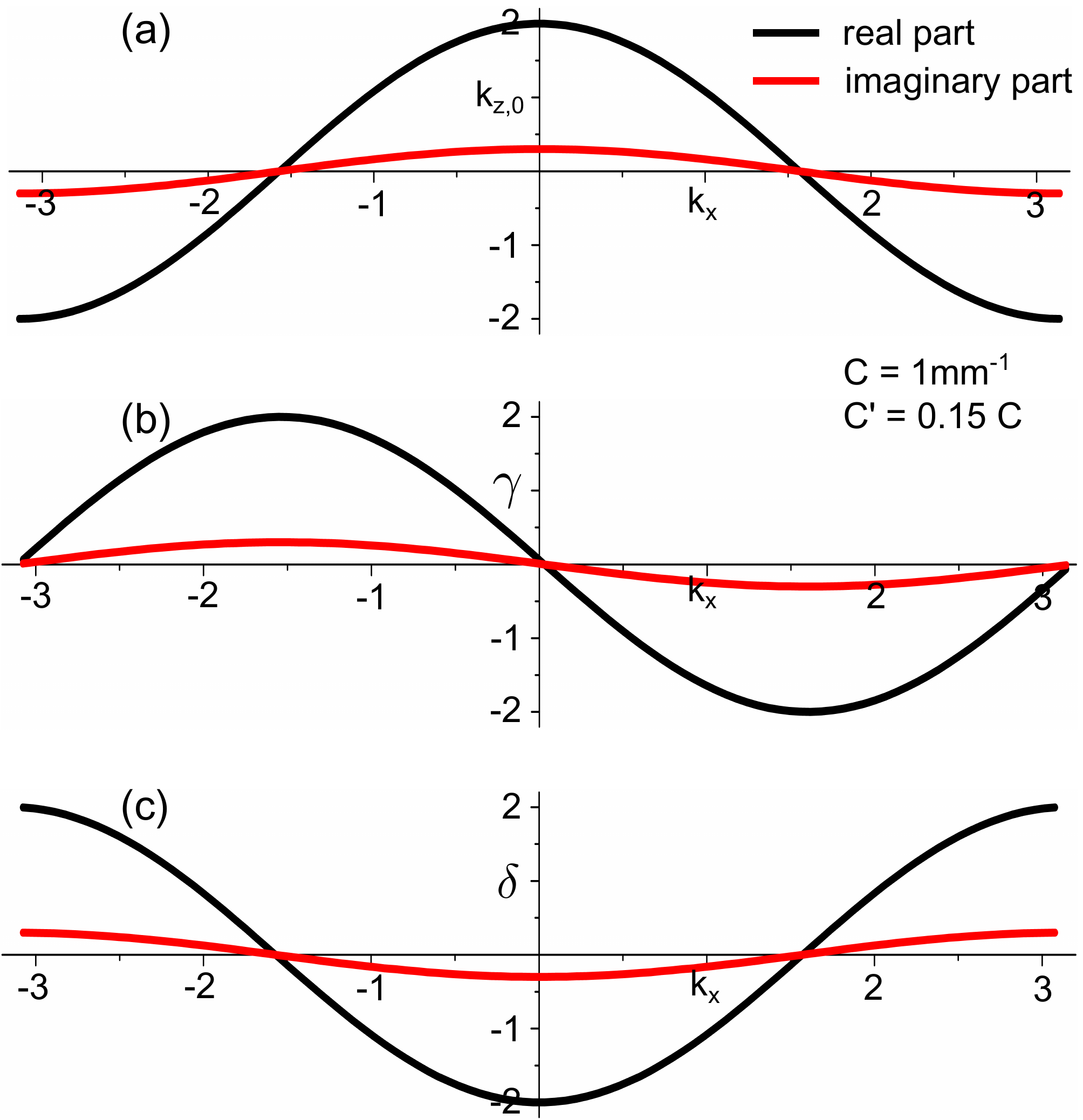}
\caption{(color online) (a) The real and imaginary parts of $k_z(k_{x,0})$. (b) The real and imaginary parts of $\gamma(k_{x,0})$. (c) The real and
imaginary parts of $\delta(k_{x,0})$.} \label{Fig3}
\end{figure}

Importantly, for any given initial condition, the field evolution is completely controlled by the dispersion relation $k_z(k_x)$. It relates the
longitudinal wave number $k_z$ to the transverse wave number $k_x$ (which we normalized by the lattice spacing) and determines how the individual
Fourier components dephase during propagation. Following the coupled mode equations for the normalized amplitudes $E_n$, the complex dispersion
relation reads as
\begin{equation}
\label{dispersionrelation}
k_z(k_x) = 2C\cos(k_x) + i2C'\cos(k_x)~.
\end{equation}
In order to study the impact of this dispersion relation on the light evolution, we follow the analysis performed in \cite{Pertsch2002} and apply it
to our complex dispersion. When a broad beam is launched into the lattice around a fixed central wave number $k_{x,0}$, the dispersion relation
\eqref{dispersionrelation} can be expanded into a Taylor series
\begin{equation}
\label{dispersionrelation_Taylor}
k_z(k_x) \approx k_{z,0} + \gamma (k_x - k_{x,0}) + \frac{\delta}{2} (k_x - k_{x,0})^2 ,
\end{equation}
with
\begin{eqnarray} \nonumber
  k_{z,0}&=& k_z(k_{x,0}) = 2C\cos(k_{x,0}) + i2C'\cos(k_{x,0}) \\
         &=& k_{z,\mathrm{r}} + ik_{z,\mathrm{i}} ~,\\[1ex] \nonumber
  \gamma &=& \left . \frac{d k_z}{d k_x} \right |_{k_{x,0}} = - 2C\sin(k_{x,0}) - i2C'\sin(k_{x,0}) \\
         &=& \gamma_{\mathrm{r}} + i\gamma_{\mathrm{i}} ~,\\[1ex] \nonumber
  \delta &=& \left . \frac{d^2 k_z}{d k_x^2} \right |_{k_{x,0}} = - 2C\cos(k_{x,0}) - i2C'\cos(k_{x,0}) \\
         &=& \delta_{\mathrm{r}} + i\delta_{\mathrm{i}} ~.
\end{eqnarray}
A plot of these quantities is shown in Fig. \ref{Fig3}. As the formal solution of Eq. \eqref{EqTightBinding} is given by Fourier decomposition,
inserting Eq. \eqref{dispersionrelation_Taylor} into this solution shows that the evolution of broad beams can be described by the partial
differential equation
\begin{equation}\label{partialdifferentialequation}
  \left \lbrack i \frac{\partial}{\partial z} - \left ( i \gamma_{\mathrm{r}}- \gamma_{\mathrm{i}} \right ) \frac{\partial}{\partial n} -
  \left ( \frac{\delta_{\mathrm{r}}}{2} + i\frac{\delta_{\mathrm{i}}}{2} \right ) \frac{\partial^2}{\partial n^2} \right \rbrack a(n,z) = 0 ~,
\end{equation}
of the distributed amplitude function
\begin{equation}
  a(n,z)= \exp \left\lbrace -i ( \lbrack k_{z,\mathrm{r}} + i k_{z,\mathrm{i}}\rbrack z + k_{x,0}n) \right\rbrace E_n(z) \; .
\end{equation}
However, it is very important to note that the validity of Eq. \eqref{partialdifferentialequation} depends strongly on the approximation of the
dispersion relation Eq. \eqref{dispersionrelation_Taylor}. If we assume that the center of mass of the normalized amplitudes $E_n(n,z)$ in the
$k_x$-space moves along $k_{x,c}(z)$ and has a variance $\Delta k_x^2(z)$, then our approximation of the dispersion relation limits the entire
analysis to cases where
\begin{align}\label{validityconditionsgeneral}
\left| {{k_{x,c}}(z)-k_{x,0}} \right| \ll 1~~~\text{and}~~~\frac{1}{{3!}}{\left( {\Delta {k_x}(z)} \right)^3} \ll 1 ~,
\end{align}
(for a detailed argumentation of these requirements see Supplemental Material).

The impact of the dispersion relation on the evolution of broad beams is best illustrated when in Eq. \eqref{partialdifferentialequation} each term
is individually analyzed, i.e., when only one quantity from the set $\lbrack \gamma_{\mathrm{r}}, \gamma_{\mathrm{i}}, \delta_{\mathrm{r}},
\delta_{\mathrm{i}}\rbrack$ is taken into account and the others are set to zero. Moreover, we would like to illustrate the new dynamics for an
initially tilted Gaussian beam
\begin{equation}\label{tiltedinputbeam-initialcondition}
  E(n,z=0)= a_0 \exp\Big(- \frac{n^2}{w_0^2}+ik_{x,0}n\Big),
\end{equation}
where $w_0$ is the initial beam width. In this case, the two conditions of Eq. (\ref{validityconditionsgeneral}) are equivalent to
\begin{equation}\label{validityconditions}
\left| {\frac{{2{\gamma _{\rm{i}}}z}}{{w_0^2 + 2{\delta _{\rm{i}}}z}}} \right| \ll 1{\mkern 1mu} {\mkern 1mu} {\mkern 1mu} \,\,\,,{\mkern 1mu}
{\mkern 1mu} \,\,\,{\mkern 1mu} {\mkern 1mu} w_0^2 + 2{\delta _{\rm{i}}}z \gg 1 \,\, .
\end{equation}
For only $\delta_{\mathrm{r}}\not=0$, these two conditions simplifies to $w_0\gg 1$, such that Eq. \ref{partialdifferentialequation} reduces to
\begin{equation}
  i \frac{\partial}{\partial z} a(n,z) =  \frac{\delta_{\mathrm{r}}}{2} \frac{\partial^2}{\partial n^2}a(n,z) \; ,
\end{equation}
which is the paraxial wave equation. Therefore, $\delta_{\mathrm{r}}$ represents the \textit{diffraction strength} and can be positive and negative,
depending on the transverse wave number $k_{x,0}$, i.e., the initial tilt of the beam. Importantly, the beam width always increases for both
$\delta_{\mathrm{r}}>0$ and $\delta_{\mathrm{r}}<0$, and stays constant for $\delta_{\mathrm{r}}=0$. The term $i \gamma_{\mathrm{r}}
\partial/\partial n$ in Eq. \ref{partialdifferentialequation} can be removed by the coordinate transformation $n \rightarrow n + \gamma_{\mathrm{r}}z$, suggesting that $\gamma_{\mathrm{r}}$
is a \textit{group velocity} (which can be also positive or negative, depending on $k_{x,0}$). This is consistent with the Hermitian case
\cite{Pertsch2002}. However, in the non-Hermitian case, there are two more quantities, $\lbrack \gamma_{\mathrm{i}}$ and
$\delta_{\mathrm{i}}\rbrack$. Interestingly, when taking into account only $\delta_{\mathrm{i}}$, then Eq. \ref{partialdifferentialequation} reduces
to
\begin{equation}
  \frac{\partial}{\partial z} a(n,z) = \frac{\delta_{\mathrm{i}}}{2} \frac{\partial^2}{\partial n^2}a(n,z) \; ,
\end{equation}
which is a diffusion equation. Therefore, the quantity $\delta_{\mathrm{i}}$ can be associated with a \textit{diffusion coefficient}. The solution of
this equation for the initial condition (\ref{tiltedinputbeam-initialcondition}) reads as
\begin{equation}
  a(n,z) = a_0 \frac{w_0}{w(z)}\exp\big(- \frac{n^2}{w^2(z)}\big) ,
\end{equation}
with the beam width
\begin{equation}
  w(z) = \sqrt{w_0^2 + 2\delta_{\mathrm{i}}z} \; .
\end{equation}
According to Eq. (\ref{validityconditions}), this result is valid if
\begin{equation}\label{27}
w_0^2 + 2{\delta _i}z \gg 1 ~.
\end{equation}
Also $\delta_{\mathrm{i}}$ can be positive or negative, however, the beam behaves differently in both cases (in contrast to $\delta_{\mathrm{r}}$).
For $\delta_{\mathrm{i}}>0$, condition \eqref{27} is always satisfied for broad input beams and one finds that
\begin{equation}
  \left . w(z) \right |_{z\rightarrow\infty} \rightarrow \sqrt{2\delta_{\mathrm{i}}z} \; ,
\end{equation}
which indeed characterizes diffusive broadening. For $\delta_{\mathrm{i}}<0$, in contrast, condition \eqref{27} is only satisfied for $z\ll w_0^2 /
2|\delta_{\mathrm{i}}|$. At larger distances, the expansion Eq. \eqref{dispersionrelation_Taylor} is not valid anymore, and standard discrete
diffraction \cite{LedererNature} dominates the light evolution. Finally, when only the quantity $\gamma_{\mathrm{i}}$ is taken into account, the
conditions for the validity of Eq. \eqref{partialdifferentialequation} are ${w_0} \gg 1$ and $z\ll w_0^2 / 2|\gamma_{\mathrm{i}}|$. The evolution of
the Gaussian input beam is then described by
\begin{equation}
\frac{\partial}{\partial z} a(n,z) =  i\gamma_{\mathrm{i}} \frac{\partial}{\partial n}a(n,z) ,
\end{equation}
yielding the solution
\begin{equation}
  a(n,z) = a_0 \exp\Big( - \frac{2i\gamma_{\mathrm{i}}zn - (\gamma_{\mathrm{i}}z)^2 + n^2}{w^2_0} \Big) \; .
\end{equation}
Hence, one can clearly see that $\gamma_{\mathrm{i}}$ causes only a deformation of the phase front but leaves the general intensity profile of the beam unchanged. Both $\delta_{\mathrm{i}}$ as well as $\gamma_{\mathrm{i}}$ are intrinsic features of the appearance of a complex coupling
coefficient. Consequently, the existence of a diffusive mobility regime does not rely on a PT-symmetric loss distribution expressed on the diagonal of the Hamiltonian~\cite{Eichelkraut2013}. Even the homogeneous loss, before thought to cause only a global, exponential decay, will eventually force the wavefunction to diffuse due to the so far not considered imaginary part of the off-diagonal elements.

\begin{figure}[t]
\includegraphics[width=\columnwidth]{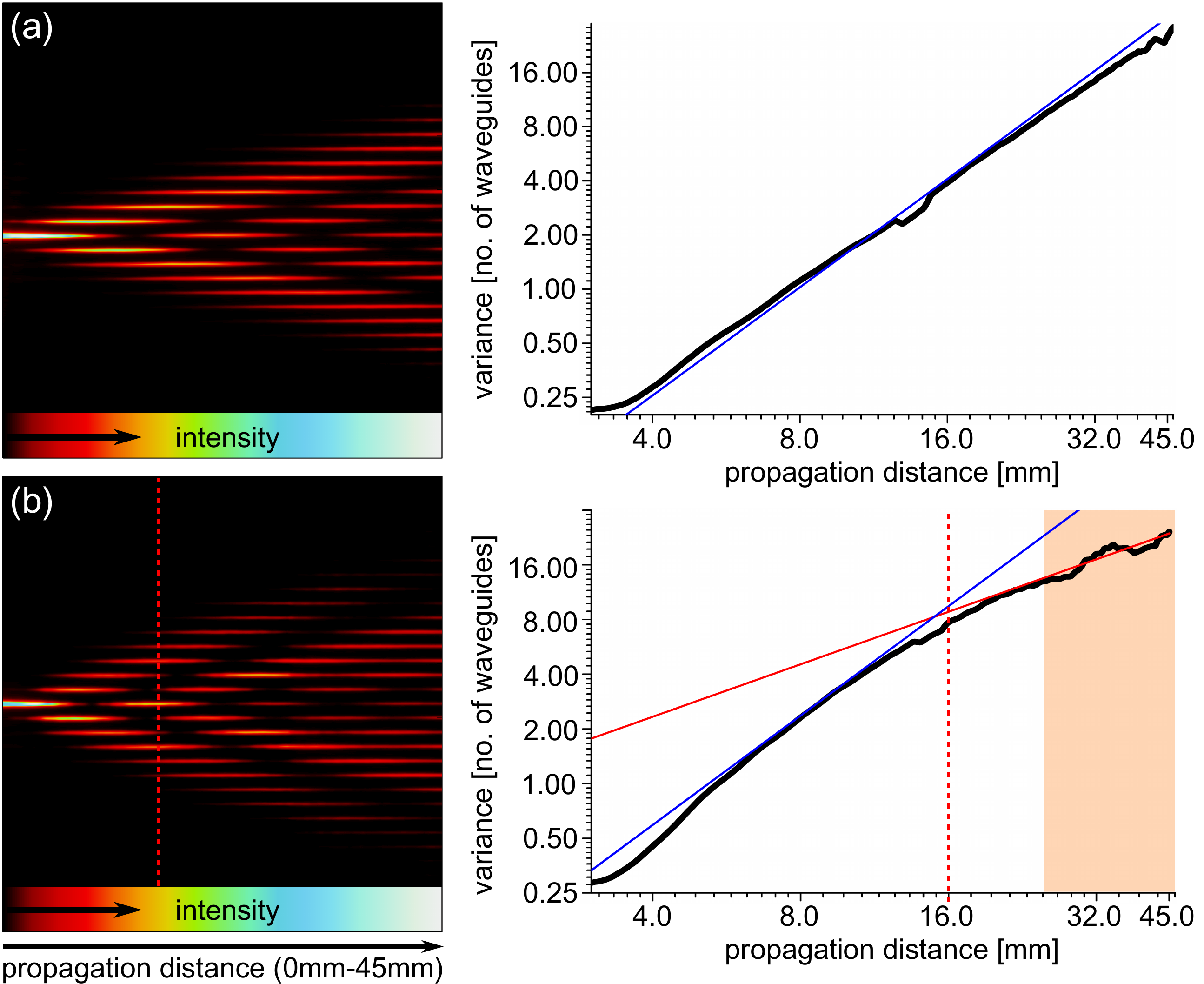}
\caption{(color online) (a) Experimental light evolution in a waveguide lattice with negligible loss (left panel). Plotting the extracted variance as
a function of the propagation distance $z$ in a double-logarithmic plot results in a straight line, which has in the ballistic case a slope of 2,
represented by the blue line (right panel). (b) Experimental light evolution in a waveguide lattice with loss $\alpha = 0.16$ (left panel). In the
double-logarithmic plot of the extracted variance one sees a transition from slope 2 (ballistic, blue line) to slope 1 (diffusive, red line). The
latter was fitted using data from the orange area, resulting in a slope of $0.96$, which is very close to the theoretical value of 1.  In boh panels, $z_{crit}$ is indicated by a red dashed line.} \label{Fig4}
\end{figure}

In order to prove the existence of the diffusive spreading in waveguide lattices with a homogeneous loss distribution (i.e., full translational
symmetry), we perform experiments in laser-written waveguide arrays in fused silica glass \cite{SzameitPhDTutorial}. For the fabrication of the
waveguides, we tightly focus ultrashort laser pulses (wavelength $515 nm$, pulse duration $308 fs$, average power $222 mW$, repetition rate $100
kHz$) using a $40\times$ objective into a 10 cm long fused silica glass wafer, which is transversely translated with $250 mm/min$ using a
high-precision positioning system. Each waveguide lattice consists of $45$ waveguides, and the spacing between the waveguides is $17 \mu m$, which
corresponds to $C = 0.1 mm^{-1}$. We analyze the light evolution in the structure by launching light at $\lambda = 633 nm$ into the central guide by
using fiber butt coupling and observe the light evolution by a fluorescence microscope technique \cite{QuasiIncoherence}. The light evolution in the
lossless array is shown in Fig. \ref{Fig4}(a), exhibiting clearly ballistic spreading. The situation changes when strong loss is introduced to the
waveguides. This is done by writing the waveguides in a sinusoidal fashion \cite{Eichelkraut2013} with an amplitude (perpendicular to $x$) of $3 \mu m$ and a period of $3 mm$, which enhances the
radiation losses of the guides. In this case it is $C' \not = 0$ and, hence, the spreading of the light field should change from ballistic to
diffusive after a particular propagation distance. This is exactly what we observe in the experiment, which is shown in Fig. \ref{Fig4} (b). The
transition occurs after $z_{\mathrm{crit}}\approx 16mm$, which implies $\alpha=0.16$ (according to Eq. \eqref{z_crit}). This is the experimental
proof that, although all waveguides exhibit the same loss, in Eq. \eqref{EqTightBinding} not only the on-diagonal loss-term $\kappa$ has to be taken
into account, but also the off-diagonal imaginary coupling $C'$ that cannot be removed by normalization.

In conclusion, we have shown that if losses are present in a photonic waveguide lattice exhibiting translational symmetry, the inter-site coupling is
complex. This results in a modified dispersion relation with an additional band due to the complex coupling. As a further consequence, the light
spreading slows down from ballistic to diffusive after a characteristic propagation distance that is determined by the loss. We believe that our
findings have fundamental impact on the understanding of light evolution in non-Hermitian lattices, in particular those with space-time reflection
($\mathcal{PT}$) symmetry~\cite{Mostafazadeh,Bender2005}. Consequently, the loss effect on transport \cite{Jovic},
which could lead to regimes such as sub/super-diffusive or even super-ballistic ones \cite{Levi2012} in addition to the changes on the band structure as a result of
higher order couplings \cite{Szameit2008} are open questions for futher investigations\cite{Eichelkraut2013}. It is also intersting to study
the impact of our results on the two-dimensional array of waveguides as candidates for ultrahigh-capacity optical communications \cite{Turitsyn2012}
or spatiotemporal vortex soliton (a result of non-linear Kerr effect) \cite{Leblond,Eilenberger2013} and their dynamical properties.

This work was supported in part by the Office of Vice-President for Research of Sharif University of Technology. A. L. gratefully acknowledges the
Alexander von Humboldt Foundation for financial support. A. S. and S. W. gratefully acknowledge financial support from the German Ministry of
Education and Research (Center for Innovation Competence program, grant 03Z1HN31), the Thuringian Ministry for Education, Science and Culture
(Research group Spacetime, grant no. 11027-514), and the Deutsche Forschungsgemeinschaft (grant NO462/6-1).


\newpage

\section{Supplemental Material}

In these supplementary notes we elaborate our calculations with some details.

\subsection{Theoretical Model}

In order to study the impact of absorption on the wave dynamics in photonic waveguide lattices,
we consider a one-dimensional array of $N$ identical single mode optical waveguides with the width $2w$ and complex
relative permittivity ${\epsilon} + i\epsilon {'}$. We conceive the waveguides written at
the positions $x_n$ ($n=1,2,...,N$) in polished fused silica bulk material with relative permittivity ${\epsilon _0} + i\epsilon {'_0}$. So, everywhere in the glass sample, the relative permittivity is considered to be a complex number; its real part corresponds to the refractive index while the imaginary part determines the absorption coefficient.

The dynamics of wave propagation through this system is governed by the Helmholtz equation
\begin{eqnarray}
  \label{EPAPSeqHelmholtz}
\left[{\nabla ^2} + k_0^2\widetilde \varepsilon (x)\right]\psi (x,z) = 0~;~~~~~~{k_0} = \frac{\omega }{c}\,,
\end{eqnarray}
where $\psi(x,z)$ is the electric field amplitude, $k_0$ is the free space propagation constant, and $\widetilde \varepsilon (x)$ is the relative permittivity profile of the entire structure. This overall permittivity distribution can be written as a sum of the individual waveguide contributions, so that
\begin{equation}
\label{EPAPSeqEpsilonProfile}
\widetilde \varepsilon (x) = {\epsilon _0} + i\epsilon {'_0} + \sum\limits_{n=1}^N \left[(\epsilon  - {\epsilon _0}) + i(\epsilon ' - \epsilon {'_0})\right]{\zeta _n}(x) ~,
\end{equation}
where ${\zeta _n}\left( x \right) = H\left( {x - {x_n} + w} \right) - H\left( {x - {x_n} - w} \right)$ equals one inside the  $n$-th guide and vanishes elsewhere (here $H\left( {x} \right)$ is the Heaviside step function). If we assume that ${\psi _s}(x,z) = u(x){e^{i\beta z}}$ is the eigenmode of one
single-mode lossless waveguide, with width $2w$, centered at $x=0$, i.e.,
$$
  {\widetilde {\varepsilon}_s}(x) = \left\{ {\begin{array}{*{20}{c}}
{\epsilon \,\,\,\,\,\,\left| x \right| < w}\\
{{\epsilon _0}\,\,\,\,\left| x \right| > w}
\end{array}} \right.
$$
one obtains the equation determining the profile of the eigenmode of a single isolated waveguide,
\begin{eqnarray}
\label{EPAPSeqHelmholtzfor1Guide}
\left( {\frac{{{d^2}}}{{d{x^2}}} + k_0^2{\widetilde {\varepsilon}_s}(x) - {\beta ^2}} \right)u(x) = 0 \,.
\end{eqnarray}
Considering now an array of waveguides according to Eq. \eqref{EPAPSeqHelmholtz}, the overall electric field $\psi$ can be expanded into
\begin{eqnarray}
\label{EPAPSeqCoupledModeApprox}
\psi (x,z) = \sum\limits_{n=1}^N {{\phi _n}(z)u(x - {x_n}){e^{i\beta z}}} .
\end{eqnarray}
The validity of this approach is limited to the coupled mode approximation. In this relation, $\beta$ is the propagation constant of the identical waveguides, $u(x-x_n)$ is the normalized transverse profile of the eigenmode of guide $n$ and ${\phi _n}(z)$ will quantify the contribution of $u(x-x_n)$ to the overall electric field. Substituting Eqs.\eqref{EPAPSeqEpsilonProfile} and
\eqref{EPAPSeqCoupledModeApprox} into Eq.\eqref{EPAPSeqHelmholtz} and using Eq.\eqref{EPAPSeqHelmholtzfor1Guide}, in the slowly varying envelope approximation, we have
\begin{widetext}
\begin{align*}
\sum\limits_{n=1}^N { - 2i\beta \left( {\frac{{d{\phi _n}(z)}}{{dz}} + \frac{{k_0^2\epsilon '}}{{2\beta }}}{\phi _n}(z) \right)} \,u(x - {x_n}) = \sum\limits_{n=1}^N {k_0^2{\phi _n}(z)u(x - {x_n})\left( {(\epsilon  - {\epsilon _0}){\Omega _n}(x) + i(\epsilon {'_0} - \epsilon ')\pi (x)} \right)} ~,
\end{align*}
\end{widetext}
where,
\begin{equation}
\label{EPAPSeqZetaandPi}
{\Omega _n}(x) = \sum\limits_{\substack{k = 1 \\ k \ne n}}^N {{\zeta _k}(x)} \,\,\,,\,\,\,\,\pi (x) = 1 - \sum\limits_{k = 1}^N {{\zeta_k}(x)}\, .
\end{equation}
By multiplying the above equation by $u(x-x_m)$, and integrating it over $x$ we obtain
\begin{equation}
\label{EPAPSeqMotionMatrixForm}
  -i\sum\limits_{n=1}^N {{V_{mn}}\left( {\frac{{d{\phi _n}(z)}}{{dz}} + \frac{{k_0^2\epsilon '}}{{2\beta }}{\phi _n}(z)} \right)}  = \sum\limits_{n=1}^N {{t_{mn}}{\phi _n}(z)} ~,
\end{equation}
with the modal overlap integrals
\begin{equation}
\label{EPAPSeqModalOverlapIntegrals}
\begin{array}{l}
{V_{mn}} = \int\limits_{ - \infty }^\infty  {u(x - {x_m})u(x - {x_n})dx} ~,
\\
\\
{t_{mn}} = {C_{mn}} + iC{'_{mn}} ~,\\
\\
{C_{mn}} = \frac{{(\epsilon  - {\epsilon _0})k_0^2}}{{2\beta }}\int\limits_{ - \infty }^\infty  {u(x - {x_m})u(x - {x_n}){\Omega _n}(x)dx} ~,\\
\\
C{'_{mn}} = \frac{{(\epsilon {'_0} - \epsilon ')k_0^2}}{{2\beta }}\int\limits_{ - \infty }^\infty  {u(x - {x_m})u(x - {x_n})\pi (x)dx} ~.
\end{array}
\end{equation}
According to the section II, if the distance between adjacent guides is much larger than the
full width half maximum (FWHM) $\ell$ of their eigenmodes, these overlap integrals can be approximated by,
\begin{equation}
\label{EPAPSeqModalIntegralinLargeDistance1}
\begin{array}{l}
{V_{mn}} \simeq {\delta _{m,n}} ~,\\
\\
{C_{mn}} \simeq {C_m}{\delta _{m,n - 1}} + {C_{m - 1}}{\delta _{m,n + 1}} ~,\\
\\
C{'_{mn}} \simeq {\kappa'}{\delta _{m,n}} + C{'_m}{\delta _{m,n - 1}} + C{'_{m - 1}}{\delta _{m,n + 1}} ~,
\end{array}
\end{equation}
with
\begin{align}
\label{EPAPSeqModalIntegralinLargeDistance2}
\kappa' &= \frac{(\epsilon {'_0} - \epsilon ')k_0^2}{2\beta}\left(1 - \text{tanh}\left(\frac{w}{\ell }\right)\right)  ~,\nonumber\\
{C_m} &= \frac{{(\epsilon  - {\epsilon _0})k_0^2}}{{2\beta }}\frac{w}{\ell }\exp \left( - \frac{{\left| {{x_{m + 1}} - {x_m}} \right|}}{\ell }\right)  ~,\\
C{'_m}& = \frac{{(\epsilon {'_0} - \epsilon ')k_0^2}}{\beta }\frac{{\left| {{x_{m + 1}} - {x_m}} \right|}}{\ell }\exp \left( - \frac{{\left| {{x_{m + 1}} - {x_m}} \right|}}{\ell }\right) . \nonumber
\end{align}
Now, the evolution equation \eqref{EPAPSeqMotionMatrixForm} simplifies considerably. To make the effect of the absorption as clear as possible, we transform out the homogeneous loss term in Eq. \eqref{EPAPSeqMotionMatrixForm}.
\begin{align}\label{EPAPSChangeofVariable}
\begin{array}{*{20}{c}}
{{\phi _n}(z) = {E_n}(z){e^{ - \kappa z}}}  ~,\\
\\
{\kappa  = \frac{{k_0^2}}{{2\beta }}\left( {\epsilon {'_0} + (\epsilon{'}- \epsilon {'_0}){\rm{tanh}}\left( {\frac{w}{\ell }} \right)} \right) .}
\end{array}
\end{align}
Here, $\kappa$ is the longitudinal attenuation factor. We finally end up with the tight-binding evolution equation
\begin{equation}\label{EPAPSEqTightBinding}
  - i\frac{{d{E_n}(z)}}{{dz}} = ({C_n} + iC{'_n}){E_{n + 1}}(z) + ({C_{n - 1}} + iC{'_{n - 1}}){E_{n - 1}}(z).
\end{equation}
One of the important features of Eq.\eqref{EPAPSEqTightBinding} is that even after the transformation \eqref{EPAPSChangeofVariable}, the effect of loss is still present in this tight-binding equation and manifests itself in complex coupling coefficients. The imaginary part of the coupling coefficients can be related to the real part by
 \begin{equation}\label{EPAPSEqImaginaryfromRealPart}
 C{'_n} = \alpha \frac{{\left| {{x_{n + 1}} - {x_n}} \right|}}{{{d_w}}}{C_n} ~,
 \end{equation}
  where we have defined
  \begin{equation}\label{EPAPSEqalpha}
 \alpha  = 2\frac{{(\epsilon {'_0} - \epsilon '){d_w}}}{{(\epsilon  - {\epsilon _0})w}}~,
  \end{equation}
the \textit{absorption discrepancy}. In this equation, $d_w = \left\langle \left| x_{n + 1} - x_n \right| \right\rangle$ is the average distance between adjacent guides.
If the absorption strength of the waveguides is different from their surrounding medium, the longitudinal attenuation of the wavepacket is not uniform in the x-direction. As we are going to show now, this trait will lead to a diffusive, transverse energy transfer.

Although the absorption discrepancy can be positive or negative, depending on the sign of $(\epsilon {'_0} - \epsilon ')$, it is easy to show that Eq. \eqref{EPAPSEqTightBinding} is invariant under transformations $\alpha\rightarrow-\alpha$ and $E_n(z) \rightarrow (-1)^n{ {E_n^{*}(z)}}$.
Therefore, in the case of single site excitation at the entrance plane where the latter transformation does not affect the initial condition, the intensity ${\left| {E_n (z)} \right|^2}$ is an even function of $\alpha$, and it is sufficient to investigate the positive values of the absorption discrepancy.

Let us finally remark that the corresponding Hamiltonian of the tight-binding Eq. \eqref{EPAPSEqTightBinding} is not a Hermitian operator and, therefore, the total power
$$\sum\limits_{n = 1}^N {{{\left| {{E_n}(z)} \right|}^2}} ~, $$ is not a conserved quantity,
\[\frac{d}{{dz}}\sum\limits_{n = 1}^N {{{\left| {{E_n}(z)} \right|}^2}}  =  - 2\sum\limits_{n = 1}^{N - 1} {C{'_n}\left( {E_{n + 1}^*{E_n} + E_n^*{E_{n + 1}}} \right)} \,.\]
Since our aim is to discuss on transverse profile, this lack of conservation is not crucial and in simulations one can normalize the total power in each $z$ step to avoid a numerical divergence.

\subsection{Estimation of Modal Overlap Integrals}

Here, we want to show the estimation of the modal overlap integrals of Eq. \eqref{eqModalOverlapIntegrals}. To this end we approximate the solution of Eq. \eqref{eqHelmholtzfor1Guide} by the following normalized function\cite{ModeofSingleWaveguide}.
\begin{equation}\label{EPAPSAP1}
u(x) = \frac{1}{{\sqrt {2\ell } }}{\mathop{\rm sech}\nolimits} (\frac{x}{\ell }) ~,
\end{equation}
where $\ell$ was the FWHM; depending on the wavelength $2\pi c/\omega$ and the waveguide parameters $\epsilon$, $\epsilon_0$ and $w$.
By using the relation
\begin{widetext}
\begin{equation}\label{EPAPSAP2}
\int {u(x - {x_m})u(x - {x_n})dx}=\frac{1}{2}{\rm{csch}}\left( {\frac{{{x_m} - {x_n}}}{\ell }} \right)\ln \left( {\frac{{\cosh\left( {\frac{{x - {x_n}}}{\ell }} \right)}}{{\cosh
\left( {\frac{{x - {x_m}}}{\ell }} \right)}}} \right) ~,
\end{equation}
\end{widetext}
\normalsize
we will find,
\begin{align*}
{V_{mn}} = {\delta _{m,n}} +&\left({1-{\delta_{m,n}}}\right) {\left| {\frac{{{x_m} - {x_n}}}{\ell }} \right|{\mathop{\rm csch}\nolimits} \left( {\left| {\frac{{{x_m} - {x_n}}}{\ell }} \right|} \right)}.
\end{align*}
Here $\delta _{m,n}$ is the Kronecker delta. In the limit of large distance between guides we have
\begin{align*}
\mathop{\rm csch}\nolimits \left(\left|\frac{x_m-x_n}{\ell }\right|\right)\approx 2\exp \left(-\left|\frac{x_m-x_n}{\ell } \right|\right).
\end{align*}
Therefore, the overlap integrals of different modes, i.e. $m \neq n$, $V_{mn}$ decay exponentially with waveguide separation. Thus, in the case of  $\left\langle {\left| {{x_{m}} - {x_n}} \right|} \right\rangle  \gg \ell $ we can write
\begin{equation}\label{EPAPSAP3}
{V_{mn}} \simeq {\delta _{m,n}}\,.
\end{equation}
Next, we will focus on the estimation of the real and imaginary part of the coupling coefficient. According to Eqs.
\eqref{EPAPSeqZetaandPi} and \eqref{EPAPSeqModalOverlapIntegrals}, for the real part of $t_{mn}$ we have
\begin{widetext}
\begin{align*}
{C_{mn}} &= \frac{{(\epsilon  - {\epsilon _0})k_0^2}}{{2\beta }}\int\limits_{ - \infty }^\infty  {u(x - {x_m})u(x - {x_n}){\Omega _n}(x)dx}= \frac{{(\epsilon  - {\epsilon _0})k_0^2}}{{2\beta }}\sum\limits_{\substack{k = 1 \\ k \ne n}}^N {\int\limits_{ - \infty }^\infty  {u(x - {x_m})u(x - {x_n}){\zeta _k}(x)dx} }\\
 &= \frac{{(\epsilon  - {\epsilon _0})k_0^2}}{{2\beta }}\sum\limits_{\substack{k = 1 \\ k \ne n}}^N {\int\limits_{{x_k} - w}^{{x_k} + w} {u(x - {x_m})u(x - {x_n})dx} }.
\end{align*}
\end{widetext}
Since $u(x - {x_n}) = \frac{1}{{\sqrt {2\ell } }}{\mathop{\rm sech}\nolimits} (\frac{{x - {x_n}}}{\ell })$ decays exponentially with $\left| x -x_n \right|$, we can approximate the above equation by the following relation
\begin{widetext}
\begin{equation*}
C_{mn} \simeq \frac{(\epsilon  - \epsilon _0)k_0^2}{2\beta } \left({\int\limits_{ - \infty }^{{x_{n - 1}} + w} {u(x - {x_m})u(x - {x_n})dx}}\right. \left. {+\int\limits_{x_{n + 1} - w}^{ \infty } u(x-x_m)u(x-x_n)dx} \right) .
\end{equation*}
\end{widetext}
By using Eq. \eqref{EPAPSAP2}, we find that
\begin{widetext}
\begin{equation*}
C_{mn} \simeq \frac{{(\varepsilon  - {\varepsilon _0})k_0^2}}{{2\beta }}\left| {\frac{{{x_m} - {x_n}}}{\ell }} \right|{\mathop{\rm csch}\nolimits} \left( {\left| {\frac{{{x_m} - {x_n}}}{\ell }}
\right|} \right) \left[1+\frac{\ell }{2(x_m - x_n)}\ln \left( {\frac{{\cosh \left( {\frac{{{x_n} - {x_{n - 1}} - w}}{\ell }} \right)\cosh \left( {\frac{{{x_{n + 1}} - {x_m} - w}}{\ell }} \right)}}{{\cosh \left( {\frac{{{x_{n + 1}}
- {x_n} - w}}{\ell }} \right)\cosh \left( {\frac{{{x_m} - {x_{n - 1}} - w}}{\ell }} \right)}}} \right) \right] .
\end{equation*}
\end{widetext}
In the case of large distances between adjacent guides, it is straightforward to show that
\begin{widetext}
\begin{align*}
C_{nn} &\simeq \frac{{(\epsilon  - {\epsilon _0})k_0^2}}{{2\beta }}\left( \exp \left(  - 2\frac{x_{n + 1} -x_n- w}{\ell } \right)+\exp \left( { - 2\frac{{{x_n} - {x_{n - 1}} - w}}{\ell }} \right) \right)~,\\
C_{n,n + 1} &= \frac{{(\epsilon  - {\epsilon _0})k_0^2}}{{2\beta }}\frac{w}{\ell }\exp \left( { - \frac{{{x_{n + 1}} - {x_n}}}{\ell }} \right)= {C_{n + 1,n}} ~,\\
{C_{mn}} &\simeq \frac{{(\epsilon  - {\epsilon _0})k_0^2}}{\beta }\frac{{\left| {{x_m} - {x_n}} \right|}}{\ell }\exp \left( { - \frac{{\left| {{x_m} - {x_n}} \right|}}{\ell }} \right)~,~~\left| {m - n} \right|\ge 2 .
\end{align*}
\end{widetext}
Comparing ${C_{mn}}$ for different values of $\left| {m - n} \right|$, we can approximate ${C_{mn}}$ by the following simple relation.
\begin{align}\label{EPAPSAP4}
{C_{mn}} &\simeq {C_m}{\delta _{m,n - 1}} + {C_{m - 1}}{\delta _{m,n + 1}} ~,\\
{C_m}& = \frac{{(\epsilon  - {\epsilon _0})k_0^2}}{{2\beta }}\frac{w}{\ell }\exp ( - \frac{{\left| {{x_{m + 1}} - {x_m}} \right|}}{\ell }) .
\end{align}
At the last step, we want to evaluate the imaginary part of the coupling coefficients, $C{'_{mn}}$. From Eqs. \eqref{EPAPSeqZetaandPi} and \eqref{EPAPSeqModalOverlapIntegrals} we have
\begin{widetext}
\begin{align*}
C{'_{mn}} = \frac{(\epsilon {'_0} - \epsilon ')k_0^2}{2\beta}&\left( \int\limits_{ - \infty }^\infty u(x - {x_m})u(x - {x_n})dx  - \sum\limits_{k=1}^N \int\limits_{x_k-w}^{x_k + w} u(x - {x_m})u(x - x_n)dx\right) .
\end{align*}
\end{widetext}
Due to the exponential decay derived in Eq. \eqref{EPAPSAP1} for $\left| {x} \right| \gg \ell$, we can approximate the last result by
\begin{widetext}
\begin{align*}
C'_{nn} \simeq \frac{(\epsilon{'_0} - \epsilon ')k_0^2}{2\beta}&\left(\int\limits_{ - \infty }^\infty u^2(x - x_n)dx-\int\limits_{x_n - w}^{x_n+w} u^2(x-x_n)dx\right),\\
C'_{n \pm 1,n} \simeq\frac{(\epsilon {'_0} - \epsilon ')k_0^2}{2\beta}&\left(\int\limits_{-\infty }^\infty u(x - x_{n \pm 1})u(x - x_n)dx \right.\\
&\left. -\int\limits_{x_{n \pm 1} - w}^{x_{n \pm 1}+w} u(x - x_{n \pm 1})u(x -x_n)dx - \int\limits_{x_n - w}^{x_n+w} u(x - x_{n \pm 1})u(x-x_n)dx \right),\\
C{'_{mn}}\simeq \frac{(\epsilon {'_0} - \epsilon ')k_0^2}{2\beta}&\int\limits_{-\infty }^\infty u(x-x_m) u(x - x_n)dx~,~~~~\left| {m - n} \right|\ge 2.
\end{align*}
\end{widetext}
These integrals can be calculated using Eq. \eqref{EPAPSAP2},
\begin{widetext}
\begin{align*}
C{'_{nn}} \simeq {\kappa'} &= \frac{{(\epsilon {'_0} - \epsilon ')k_0^2}}{{2\beta }}\left(1-\tanh \left( \frac{w}{\ell}\right)\right)~,\\
C{'_{n,n + 1}} &\simeq \frac{{(\epsilon {'_0} - \epsilon ')k_0^2}}{{2\beta }}{\mathop{\rm csch}\nolimits} \left( {\frac{{{x_{n + 1}} - {x_n}}}{\ell }} \right)\left( \left( {\frac{{{x_{n + 1}} - {x_n}}}{\ell }} \right) - \ln \left( {\frac{{\cosh \left( {\frac{{{x_{n + 1}} - {x_n} + w}}{\ell }}\right)}}{{\cosh \left( {\frac{{{x_{n + !}} - {x_n} - w}}{\ell }} \right)}}} \right) \right) = C{'_{n + 1,n}} ~,\\
C{'_{mn}}&\simeq \frac{{(\epsilon {'_0} - \epsilon ')k_0^2}}{{2\beta }}\left| {\frac{{{x_m}-{x_n}}}{\ell }} \right|{\mathop{\rm csch}\nolimits} \left( {\left| {\frac{{{x_m} - {x_n}}}{\ell }} \right|} \right)~,~~\left| {m - n}\right|\ge 2 .
\end{align*}
\end{widetext}
By expanding the above relations at the limit of large distance between neighboring guides, for $\left| {m - n} \right| \ge 1$, we have
$$C{'_{mn}} = \frac{{(\epsilon {'_0} - \epsilon ')k_0^2}}{{2\beta }}\left| {\frac{{{x_m} - {x_n}}}{\ell }} \right|\exp \left( { - \left| {\frac{{{x_m} - {x_n}}}{\ell }} \right|} \right).$$
Therefore, in the limit of $\left\langle {\left| {{x_{n+1}} - {x_n}} \right|} \right\rangle  \gg \ell $, we can write
\begin{equation}\label{EPAPSAP5}
{C{'_{mn}} \simeq {\kappa'}{\delta _{m,n}} + C{'_m}{\delta _{m,n - 1}} + C{'_{m - 1}}{\delta _{m,n + 1}}}~,
\end{equation}
where
\begin{align}\label{EPAPSAP6}
{\kappa'} &= \frac{{(\epsilon {'_0} - \epsilon ')k_0^2}}{{2\beta }}\left( {1 - \tanh \left( {\frac{w}{\ell }} \right)} \right)~,\\
C{'_m} &= \frac{{(\epsilon {'_0} - \epsilon ')k_0^2}}{\beta }\frac{{\left| {{x_{m + 1}} - {x_m}} \right|}}{\ell }\exp ( - \frac{{\left| {{x_{m + 1}} - {x_m}} \right|}}{\ell }) .
\end{align}

\subsection{Variance}

According to section I, in a one-dimensional periodic array of identical lossy waveguides the electric field amplitude at each waveguide will
satisfy the following tight-binding equation,
\begin{widetext}
\begin{equation}\label{EPAPSEqTgihtBindingPeriodic}
{\phi _n}(z) = {E_n}(z){e^{ - \kappa z}}\,\,,\,\,\, - i\frac{{d{E_n}(z)}}{{dz}} = C\left( {1 + i\alpha } \right)\left( {{E_{n + 1}}\left( z \right) +
{E_{n - 1}}\left( z \right)} \right).
\end{equation}
\end{widetext}
Again, $\kappa$, $C$ and $\alpha$ are the attenuation constant along the propagation direction $z$, the real part of the coupling coefficient and the absorption discrepancy, respectively.\\
For single-site excitation, $E_n(0)=\delta_{n,n_0}$, the solution of this equation is \cite{PRA2008-LongrangeInteraction}
\begin{equation}\label{EPAPSPeriodicSolutionJn}
{E_n}(z) = {i^{n - {n_0}}}{J_{n - {n_0}}}\left( {2\left( {1 + i\alpha } \right)Cz} \right),
\end{equation}
where $J_n(x)$ represents the Bessel function of order n.

Starting from Eq. \eqref{EPAPSEqTgihtBindingPeriodic}, the eigenvalue spectrum of this system of equations can be obtained using the plane wave approach $E_n^q(z)
= {E_0}\left( q \right){e^{i\left( {\beta \left( q \right)z + qn} \right)}}$. The dispersion relation reads
\begin{equation}\label{EPAPSDispersionRelation}
\beta \left( q \right) = \left( {1 + i\alpha } \right){\beta _r}\left( q \right)\,\,\,,\,\,\,\,{\beta _r}\left( q \right) = 2C\cos \left( q \right).
\end{equation}
In this case both the real part and imaginary part of the propagation constant $\beta \left( q \right)$ are dispersive. Using the initial condition
${E_n}\left( {z = 0} \right) = {\delta _{n,{n_0}}}$, the evolution of the wavepacket can be written as
\begin{equation}\label{EPAPSWavepacketEvolution}
{E_n}\left( z \right) = \frac{1}{{2\pi }}\int\limits_{ - \pi }^\pi  {dq{e^{ - \left( {\alpha  - i} \right){\beta _r}\left( q \right)z}}{e^{iq\left(
{n - {n_0}} \right)}}},
\end{equation}
and therefore,
\begin{widetext}
\begin{equation}\label{EPAPSWavepacketIntensity}
\left| {{E_n}\left( z \right)} \right|^2 = \frac{1}{{4{\pi ^2}}}\int\limits_{ - \pi }^\pi  {d{q_1}\int\limits_{ - \pi }^\pi  {d{q_2}{e^{2iCz\left(
{\cos \left( {{q_2}} \right) - \cos \left( {{q_1}} \right)} \right)}}{e^{ - 2\alpha Cz\left( {\cos \left( {{q_2}} \right) + \cos \left( {{q_1}}
\right)} \right)}}{e^{i\left( {{q_2} - {q_1}} \right)\left( {n - {n_0}} \right)}}} } \ .
\end{equation}
\end{widetext}
The variance of the wavepacket is commonly defined as the following
\begin{equation}\label{EPAPSVarianceDefinition}
{\sigma ^2}\left( z \right) = \frac{{\sum\limits_{n =  - \infty }^\infty  {{{\left( {n - {n_0}} \right)}^2}{{\left| {{E_n}\left( z \right)}
\right|}^2}} }}{{\sum\limits_{n =  - \infty }^\infty  {{{\left| {{E_n}\left( z \right)} \right|}^2}} }} = \frac{{\sum\limits_{n =  - \infty }^\infty
{{n^2}{{\left| {{E_{n + {n_0}}}\left( z \right)} \right|}^2}} }}{{\sum\limits_{n =  - \infty }^\infty  {{{\left| {{E_{n + {n_0}}}\left( z \right)}
\right|}^2}} }} \ .
\end{equation}
Using Eq. \eqref{EPAPSWavepacketIntensity}, we have
\begin{widetext}
$$\sum\limits_{n =  - \infty }^\infty  {{{\left| {{E_{n + {n_0}}}\left( z \right)} \right|}^2}}  = \frac{1}{{4{\pi ^2}}}\int\limits_{ - \pi }^\pi  {d{q_1}\int\limits_{ - \pi }^\pi  {d{q_2}{e^{2iCz\left( {\cos \left( {{q_2}} \right) - \cos \left( {{q_1}} \right)} \right)}}{e^{ - 2\alpha Cz\left( {\cos \left( {{q_2}} \right) + \cos \left( {{q_1}} \right)} \right)}}\left( {\sum\limits_{n =  - \infty }^\infty  {{e^{i\left( {{q_2} - {q_1}} \right)n}}} } \right)} } .$$
\end{widetext}
However, for $\left| z \right| < 2\pi$,
\begin{equation}\label{EPAPSDeltaandIts2ndDerivative}
\sum\limits_{n =  - \infty }^\infty  {{e^{izn}}}  = 2\pi \delta \left( z \right)\,\,\,\,,\,\,\,\,\,\,\sum\limits_{n =  - \infty }^\infty
{{n^2}{e^{izn}}}  =  - 2\pi \frac{{{\partial ^2}}}{{\partial {z^2}}}\delta \left( z \right) ~,
\end{equation}
which leads to,
\begin{equation}\label{EPAPSNumeratorofVariance}
\sum\limits_{n =  - \infty }^\infty  {{{\left| {{E_n}\left( z \right)} \right|}^2}}  = \frac{1}{{2\pi }}\int\limits_{ - \pi }^\pi  {d{q_1}{e^{ -
4\alpha Cz\cos \left( {{q_1}} \right)}}}  = {I_0}\left( {4\alpha Cz} \right),
\end{equation}
where $I_0(x)$ is the zero order modified Bessel function of the first kind. Moreover,
\begin{widetext}
$$\sum\limits_{n =  - \infty }^\infty  {{n^2}{{\left| {{E_{n + {n_0}}}\left( z \right)} \right|}^2}}  = \frac{1}{{4{\pi ^2}}}\int\limits_{ - \pi }^\pi  {d{q_1}\int\limits_{ - \pi }^\pi  {d{q_2}{e^{2iCz\left( {\cos \left( {{q_2}} \right) - \cos \left( {{q_1}} \right)} \right)}}{e^{ - 2\alpha Cz\left( {\cos \left( {{q_2}} \right) + \cos \left( {{q_1}} \right)} \right)}}\left( {\sum\limits_{n =  - \infty }^\infty  {{n^2}{e^{i\left( {{q_2} - {q_1}} \right)n}}} } \right)} } \ .$$
\end{widetext}
Using the second relation of Eq. \eqref{EPAPSDeltaandIts2ndDerivative} and the fact that,
$$\int {f\left( z \right)} \frac{{{\partial ^2}}}{{\partial {z^2}}}\delta \left( {z - {z_0}} \right)dz = {\left. {\frac{{{\partial ^2}}}{{\partial {z^2}}}f\left( z \right)} \right|_{z = {z_0}}} \ ,$$
we find
\begin{widetext}
$$\sum\limits_{n =  - \infty }^\infty  {{n^2}{{\left| {{E_{n + {n_0}}}\left( z \right)} \right|}^2}}  = \frac{{ - 1}}{{2\pi }}2Cz\left( {\alpha  - i} \right)\int\limits_{ - \pi }^\pi  {d{q_1}{e^{ - 4\alpha Cz\cos \left( {{q_1}} \right)}}\left( {\cos \left( {{q_1}} \right) + 2Cz\left( {\alpha  - i} \right){{\sin }^2}\left( {{q_1}z} \right)} \right)} ~,$$
\end{widetext}
which can be calculated analytically.
\begin{equation}\label{EPAPSDenominatorofVariance}
\sum\limits_{n =  - \infty }^\infty  {{n^2}{{\left| {{E_{n + {n_0}}}\left( z \right)} \right|}^2}}  = \left( {\alpha  + \frac{1}{\alpha }}
\right)Cz\,{I_1}\left( {4\alpha Cz} \right) \ .
\end{equation}
Here, $I_1(x)$ is the first order modified Bessel function of the first kind. Combining Eqs. \eqref{EPAPSVarianceDefinition}, \eqref{EPAPSNumeratorofVariance}
and \eqref{EPAPSDenominatorofVariance}, we obtain the exact relation for the variance of the wavepacket at any value of propagation distance $z$,
\begin{equation}\label{EPAPSVarianceofWavepacket}
\sigma ^2\left( z \right) = \left( {\alpha  + \frac{1}{\alpha }} \right)Cz\,\frac{{{I_1}\left( {4\alpha Cz} \right)}}{I_0\left(
4\alpha Cz \right)}\ .
\end{equation}
Using the asymptotic expansion of the modified Bessel functions at small and large values of its argument \cite{Arfken-MathematicalPhysics},
\begin{align*}
I_\nu\left( x \right)&\mathop  \to \limits^{x \ll 1} \frac{1}{{\nu !}}{\left( {\frac{x}{2}} \right)^\nu } +  \cdots ~,\\
I_\nu\left( x \right)&\mathop  \to \limits^{x \gg 1} \frac{{{e^x}}}{{\sqrt {2\pi x} }} +  \cdots ~,
\end{align*}
we obtain
\begin{align}\label{EPAPSAssymptoticVarianceofWavepacket}
\sigma ^2\left( z \right)&\mathop  \to \limits^{4\alpha Cz \ll 1}2\left( {1 + {\alpha ^2}} \right){C^2} z^2 ~,\nonumber\\
\sigma ^2\left( z \right)&\mathop  \to \limits^{4\alpha Cz \gg 1}\left( {\frac{{1 + {\alpha ^2}}}{\alpha }} \right)C z ~,\\
{\sigma ^2}\left( z \right)&\mathop  \to \limits^{\alpha  \to 0} 2{C^2} z^2 ~, \nonumber
\end{align}
which shows that, in the regime of small values of normalized propagation distances $Cz$, or in the absence of absorption, the propagation is ballistic. On the other hand, for the large values of $Cz$ and $\alpha \neq 0$ the regime is diffusive. Moreover, using these relations, we can approximate the propagation distance at which the transfer regime changes from ballistic to diffusive.
\begin{equation}\label{EPAPSCriticalZ}
{z_{crit}} = \frac{1}{{4\alpha C}} .
\end{equation}

\subsection{Broad beam propagation in lossy photonic waveguide lattices}


Consider a one dimensional periodic array of identical lossy waveguides. The electric filed amplitude at each waveguide will satisfy Eq. (\ref{EqTgihtBindingPeriodic}). A plane wave expansion solution of Eq. (\ref{EqTgihtBindingPeriodic}) can be obtained as
\begin{equation}\label{EPAPSplanewavesolution}
\begin{array}{l}
{E_n}(z) = \int\limits_{ - \pi }^\pi  {d{k_x}\tilde E\left( {{k_x}} \right)\exp \left( {i\left( {{k_z}\left( {{k_x}} \right)z + {k_x}n} \right)} \right)} ~,
\end{array}
\end{equation}
with the discrete Fourier amplitude
\begin{equation}\label{EPAPSdiscretefouriertransform}
\tilde E\left( {{k_x}} \right) = \frac{1}{{2\pi }}\sum\limits_n {{E_n}(z = 0)\exp \left( { - i{k_x}n} \right)} ~,
\end{equation}
and the complex dispersion relation
\begin{equation}\label{EPAPSdispersionrelation}
\begin{array}{l}
{k_z}\left( {{k_x}} \right) = 2C\cos \left( {{k_x}} \right) + i2C'\cos \left( {{k_x}} \right),
\end{array}
\end{equation}
which relates the longitudinal wave number $k_z$ to the transverse wave number $k_x$. Here, $C'=\alpha C$ is the imaginary part of the coupling coefficient.
\\
When a broad beam is launched into the lattice around a fixed central wavenumber $k_{x,0}$, i.e. if
${\left| {\tilde E\left( {{k_x}} \right)} \right|^2}$ centered around $k_{x,0}$ has a small variance $(\Delta k_x)^2$, the dispersion relation (\ref{EPAPSdispersionrelation}) can be expanded into a Taylor series
\begin{equation}\label{EPAPSdispersionrelation_Taylor}
{k_z}\left( {{k_x}} \right) \approx {k_{z,0}} + \gamma \left( {{k_x} - {k_{x,0}}} \right) + \frac{\delta }{2}{\left( {{k_x} - {k_{x,0}}} \right)^2} ~,\,
\end{equation}
with
\begin{widetext}
\begin{eqnarray}
  k_{z,0}&=& k_z(k_{x,0}) = 2C\cos(k_{x,0}) + i2C'\cos(k_{x,0}) =k_{z,\mathrm{r}} + ik_{z,\mathrm{i}} ~,\\[1ex]
  \gamma &=& \left . \frac{d k_z}{d k_x} \right |_{k_{x,0}} = - 2C\sin(k_{x,0}) - i2C'\sin(k_{x,0})=\gamma_{\mathrm{r}} + i\gamma_{\mathrm{i}} ~,\\[1ex]
  \delta &=& \left . \frac{d^2 k_z}{d k_x^2} \right |_{k_{x,0}} = - 2C\cos(k_{x,0}) - i2C'\cos(k_{x,0})=\delta_{\mathrm{r}} + i\delta_{\mathrm{i}} ~.
\end{eqnarray}
\end{widetext}
Inserting Eq. (\ref{EPAPSdispersionrelation_Taylor}) into Eq. (\ref{EPAPSplanewavesolution}), the distributed amplitude function
\begin{equation}\label{EPAPSrelationanEn}
a\left( {n,z} \right) = {E_n}(z)exp\left( { - i{[k_{z,\mathrm{r}} + ik_{z,\mathrm{i}}]}z - i{k_{x,0}}n} \right) \, ,
\end{equation}
can be represented in the integral form
\begin{widetext}
\begin{equation}\label{EPAPSplanewavesolution-anz}
a\left( {n,z} \right) = \int\limits_{ - \pi  - {k_{x,0}}}^{\pi  - {k_{x,0}}} {d{k_x}\tilde E\left( {{k_{x,0}} + {k_x}} \right)\exp \left\{ {i{k_x}\left( {n + \gamma z} \right) + i\frac{{\delta z}}{2}{k_x}^2} \right\}} \,,
\end{equation}
\end{widetext}
and therefore, the evolution of $a\left( {n,z} \right)$ can be described by the following partial differential equation
\begin{equation}\label{EPAPSpartialdifferentialequation}
  \left \lbrack i \frac{\partial}{\partial z} - \left ( i \gamma_{\mathrm{r}}- \gamma_{\mathrm{i}} \right ) \frac{\partial}{\partial n} -
  \left ( \frac{\delta_{\mathrm{r}}}{2} + i\frac{\delta_{\mathrm{i}}}{2} \right ) \frac{\partial^2}{\partial n^2} \right \rbrack a(n,z) = 0 \, .
\end{equation}
It is necessary to note that the Taylor series \eqref{EPAPSdispersionrelation_Taylor}, and therefore Eqs. \eqref{EPAPSplanewavesolution-anz} and \eqref{EPAPSpartialdifferentialequation}, are only valid for small values of $\,\left| {{k_x} - {k_{x,0}}} \right|$, such that the reminder term of the Taylor expansion of the $cosine$ function is very small. Since $\left| {\sin \left( {{k_x}} \right)} \right| \le 1$, this condition reads as \cite{Arfken-MathematicalPhysics}
\begin{equation}\label{EPAPSvalidityofTaylorcondition}
\frac{1}{{3!}}{\left| {{k_x} - {k_{x,0}}} \right|^3} \ll 1 \, .
\end{equation}
However, in Eq. \eqref{EPAPSplanewavesolution-anz} the integrand will in general be evaluated beyond the region where Eq. \eqref{EPAPSvalidityofTaylorcondition} holds. Since only the variance $(\Delta k_x)^2$ of $\tilde E\left( {{k_{x,0}} + {k_x}} \right)$ can confine the actual region of integration, we impose that $\left| {{k_x} - {k_{x,0}}} \right|\approx\Delta k_x$ must satisfy condition \eqref{EPAPSvalidityofTaylorcondition} in order to ensure the validity of Eqs. \eqref{EPAPSplanewavesolution-anz} and \eqref{EPAPSpartialdifferentialequation}. So, for input beams that are broad enough in $n$-space we can write
\begin{equation}\label{EPAPSplanewavesolution-anzinfinity}
{a }\left( {n,z} \right) = \int\limits_{ - \infty }^\infty  {d{k_x}\tilde E\left( {{k_{x,0}} + {k_x}} \right)\exp \left\{ {i{k_x}\left( {n + \gamma z} \right) + i\frac{{\delta z}}{2}{k_x}^2} \right\}}\, .
\end{equation}
In the next step we define the continuous Fourier transform
\begin{equation}\label{EPAPScontinousfouriertransform}
\begin{array}{*{20}{c}}
{\tilde a\left( {{k_x},z} \right) = \frac{1}{{2\pi }}\int\limits_{ - \infty }^\infty  {dn\,a\left( {n,z} \right)\exp \left( { - i{k_x}n} \right)} }~,\\
\\
{a\left( {n,z} \right) = \int\limits_{ - \infty }^\infty  {d{k_x}\,\tilde a\left( {{k_x},z} \right)\exp \left( {i{k_x}n} \right)} }~.
\end{array}
\end{equation}
Implementing equations (\ref{EPAPSplanewavesolution-anzinfinity}) and (\ref{EPAPScontinousfouriertransform}), we obtain
\begin{equation}\label{EPAPSfouriertransform-anzinfinity}
{{\tilde a} }\left( {{k_x},z} \right) = \tilde E\left( {{k_{x,0}} + {k_x}} \right)\exp \left\{ {i{k_x}\gamma z + i\frac{{\delta z}}{2}{k_x}^2} \right\}\, .
\end{equation}
For the tilted input Gaussian beam (with angle $\theta$ and wavelength $\lambda$)
\begin{equation}\label{EPAPSinitialcondition-realEn}
{E_n}(z = 0) = {a_0}\exp \left( { - \frac{{{n^2}}}{{w_0^2}} + i{k_{x,0}}x} \right)\,\,\,\,,\,\,\,{k_{x,0}} = \frac{{2\pi d}}{\lambda }\sin \left( \theta  \right) ~,
\end{equation}
i.e.,
\begin{equation}\label{EPAPSinitialcondition-real}
a(n,z = 0) = {a_0}\exp \left( { - \frac{{{n^2}}}{{w_0^2}}} \right) .
\end{equation}
From Eq. (\ref{EPAPScontinousfouriertransform}), one gets
\begin{equation}\label{EPAPSinitialcondition-fourier}
\tilde a\left( {{k_x},z = 0} \right)=\tilde E\left( {{k_{x,0}} + {k_x}} \right) = \frac{{{a_0}{w_0}}}{{2\sqrt \pi  }}\exp \left( { - \frac{{w_0^2{k_x}^2}}{4}} \right) \, .
\end{equation}
Here, we must note that although the discrete Fourier transform \eqref{EPAPSdiscretefouriertransform} is the true equation for obtaining $\tilde a\left( {{k_x},z = 0} \right)=\tilde E\left( {{k_{x,0}} + {k_x}} \right)$, for the broad input beam, we can use the continuous Fourier transform \eqref{EPAPScontinousfouriertransform} as well.\\
Substituting  Eq. (\ref{EPAPSinitialcondition-fourier}) in Eq. (\ref{EPAPSfouriertransform-anzinfinity}), we find
\begin{equation}\label{EPAPSfourieramplitudeanz}
\tilde a\left( {{k_x},z} \right) = \frac{{{a_0}{w_0}}}{{2\sqrt \pi  }}\exp \left( { - \frac{{w_0^2{k_x}^2}}{4}} \right)\exp \left\{ {i{k_x}\gamma z + i\frac{{\delta z}}{2}{k_x}^2} \right\} \, ,
\end{equation}
which can be brought into the following intuitive expression,
\begin{widetext}
\begin{equation}\label{EPAPSfourieramplitudeanz-completesquare}
\tilde a\left( {{k_x},z} \right) = \frac{{{a_0}{w_0}}}{{2\sqrt \pi  }}exp\left\{ {\frac{{{\gamma _{\rm{i}}}^2{z^2}}}{{w_0^2 + 2{\delta _{\rm{i}}}z}}} \right\}\exp \left\{ { - \left( {\frac{{w_0^2 + 2{\delta _{\rm{i}}}z}}{4}} \right){{\left( {{k_x} + \frac{{2{\gamma _{\rm{i}}}z}}{{w_0^2 + 2{\delta _{\rm{i}}}z}}} \right)}^2}} \right\}\exp \left\{ {i{k_x}{\gamma _{\rm{r}}}z + i\frac{{{\delta _{\rm{r}}}z}}{2}{k_x}^2} \right\} .
\end{equation}
\end{widetext}
This equation and Eq. (\ref{EPAPSrelationanEn}) state that the Fourier transform of $E_n(z)$ (i.e. $\tilde E\left( {{k_{x,}} ,z} \right)$) is a Gaussian function with the center
\begin{equation}\label{EPAPScenterfourier}
{k_{x,c}}\left( z \right) =  k_{x,0}- \frac{{2{\gamma _{\rm{i}}}z}}{{w_0^2 + 2{\delta _{\rm{i}}}z}} ~,
\end{equation}
and the variance
\begin{equation}\label{EPAPSwidthfourier}
(\Delta {k_x})^2\left( z \right) = \frac{2}{{w_0^2 + 2{\delta _{\rm{i}}}z}}\, .
\end{equation}
All in all, we find that Eq. (\ref{EPAPSpartialdifferentialequation}) is valid whenever
(i) the center ${k_{x,c}}\left( z \right)$ is not very different from $k_{x,0}$ and
(ii) the variance $\Delta {k_x}^2\left( z \right)$ satisfies the condition (\ref{EPAPSvalidityofTaylorcondition}), i.e.
\begin{equation}\label{EPAPSvalidityofTaylorconditionintermofParameters}
\left| {\frac{{2{\gamma _{\rm{i}}}z}}{{w_0^2 + 2{\delta _{\rm{i}}}z}}} \right| \ll 1{\mkern 1mu} {\mkern 1mu} {\mkern 1mu} \,\,\,,{\mkern 1mu} {\mkern 1mu} \,\,\,{\mkern 1mu} {\mkern 1mu} w_0^2 + 2{\delta _{\rm{i}}}z \gg 1 \,\, .
\end{equation}


\end{document}